\documentclass[aps,
                prb,
                reprint,
                showkeys, 
                superscriptaddress, 
                nofootinbib,
                a4paper
                ]{revtex4-2}

\usepackage{amsfonts}
\usepackage{balance}
\usepackage{amsmath}
\usepackage{amssymb}
\usepackage{graphicx}
\usepackage{adjustbox}
\usepackage{circuitikz}
\usepackage{import}
\usepackage{subfigure}
\usepackage{physics}
\usepackage{adjustbox}
\usepackage{natbib}

\usepackage{bm}
\usepackage{mathtools}

\usepackage{xcolor}
\definecolor{ceruleanblue}{rgb}{0.16, 0.32, 0.75}
\definecolor{C0}{rgb}{0.12156862745098039, 0.4666666666666667, 0.7058823529411765}
\definecolor{C3}{rgb}{0.8392156862745098, 0.15294117647058825, 0.1568627450980392}

\definecolor{pastelgreen}{rgb}{0.47, 0.87, 0.47}
\definecolor{pastelblue}{rgb}{0.68, 0.78, 0.81}
\definecolor{pastelorange}{rgb}{1.0, 0.7, 0.28}
\definecolor{pastelpurple}{rgb}{0.7, 0.62, 0.71}

\definecolor{wongblue}{rgb}{0.0,0.44705883,0.69803923}
\definecolor{wongyellow}{rgb}{0.9019608,0.62352943,0.0}
\definecolor{wonggreen}{rgb}{0.0,0.61960787,0.4509804}
\definecolor{wongpink}{rgb}{0.8,0.4745098,0.654902}
\definecolor{wonglightblue}{rgb}{0.3372549,0.7058824,0.9137255}
\definecolor{wongorange}{rgb}{0.8352941,0.36862746,0.0}
\definecolor{wonglightyellow}{rgb}{0.9411765,0.89411765,0.25882354}

\usepackage[colorlinks=true,
			allcolors=ceruleanblue]{hyperref}

\usepackage{orcidlink}

\usepackage{parskip}

\usepackage{tikz}
\usetikzlibrary{arrows.meta, positioning, shapes.geometric}
\usetikzlibrary{decorations.text}
\usetikzlibrary {arrows.meta} 
\usetikzlibrary{backgrounds}

\usepackage{newtxtext,newtxmath}
\usepackage{microtype}

\newcommand{\imu}{\mathrm{i}}

\newcommand{\imec}{Imec, Kapeldreef 75, 3001 Heverlee, Belgium}
\newcommand{\itf}{Instituut voor Theoretische Fysica, KU Leuven, Celestijnenlaan 200D, 3001 Heverlee, Belgium}
\newcommand{\esat}{Department of Electrical Engineering, KU Leuven, Kasteelpark Arenberg 10, 3001 Heverlee, Belgium}
\newcommand{\ua}{Department of Physics, University of Antwerp, Groenenborgerlaan 171, 2020 Antwerp, Belgium}

\newcommand{\titlex}{Current Conservation in the Self-Consistent Josephson Junction}

\hypersetup{
	pdfauthor={Krekels et al.},
	pdftitle={\titlex}}
 
\begin{document}

\title{\titlex}

\author{Simon Krekels\,\orcidlink{0000-0003-1726-6625}}\email{simon.krekels@imec.be}\affiliation{\imec}\affiliation{\itf}
\author{Vukan Levajac\,\orcidlink{0000-0002-6985-822X}}\affiliation{\imec}
\author{Kristof Moors\,\orcidlink{0000-0002-8682-5286}}\affiliation{\imec}
\author{George Simion\,\orcidlink{0000-0002-6880-6161}}\affiliation{\imec}
\author{Bart Sor\'ee\,\orcidlink{0000-0002-4157-1956}}\affiliation{\imec}\affiliation{\esat}\affiliation{\ua}
\date{\today}

\begin{abstract}
    Conventional treatments of Josephson junctions (JJs) are typically not current-conserving.
    In the mean-field BCS theory, current conservation is only guaranteed if the superconducting order parameter is treated self-consistently.
    We show that this requirement has significant consequences for the current-phase relation (CPR) in certain regimes, where the current density in the superconducting leads is non-negligible.
    To this end, we introduce a numerical method for the self-consistent treatment of the BdG equations with current conservation for quasi-1D superconductor-normal (metal)-superconductor (SNS) JJs.
    Our model incorporates a phase gradient of the order parameter in the leads, which is set to match the Josephson current through the weak link.
    We compare our method to standard, non-current-conserving approaches by calculating the CPR for SNS JJs while varying lengths and gate voltages controlling the normal metal.
    We show that current conservation has significant implications for the Josephson harmonics and can weaken or even reverse forward skewness of the CPR. 
\end{abstract}

\maketitle

\section{Introduction}
The dissipationless current through a Josephson Junction (JJ) is its fundamental property, with the current amplitude and direction dependent on the superconducting phase difference between the superconductors \cite{josephson_possible_1962, josephson_supercurrents_1965}.
This phase-dependent Josephson current has been widely exploited in numerous applications in which JJs are used as building blocks for superconducting qubits or superconducting quantum interference devices (SQUIDs) \cite{benz2019Josephson,makhlin2001Quantumstate,kjaergaard_superconducting_2020,martinis2020quantum,tolpygo2016superconductor}.
Since the precise shape of the Josephson current's current-phase relation (CPR) determines the anharmonicity of qubits and phase sensitivity of SQUIDs, solid understanding and precise control of the CPR are essential for these applications \cite{blais_circuit_2021}.
Despite this high practical importance of the CPR, a theoretical description of JJs from a microscopic perspective remains challenging \cite{miller2001Microscopic,choi2022Microscopic}.
The need for reliable microscopic descriptions of the CPR is strengthened by the fact that CPRs, which in many cases are sinusoidal, can exhibit a particular skewness and even evolve to more exotic shapes \cite{golubov_current-phase_2004,dellarocca2007Measurement,spanton2017Current,levajac2024Supercurrenta}.
Therefore, novel theoretical methods are necessary to improve the understanding of various features of the CPR, as fabrication and measurement techniques are rapidly advancing \cite{spanton2017Current,levajac2024Supercurrenta,chen2024Current,nanda2017CurrentPhase,babich2023Limitations,vandamme2024Advanced,kringhoj2018Anharmonicity, willsch_observation_2024}.


By modeling JJs in the BCS theory of superconductivity, one can gain access to the full quasiparticle excitation spectrum of the system.
Because the JJ setup is naturally spatially inhomogeneous, the system should be treated using Bogoliubov-de Gennes (BdG) theory, or the Gor'kov Green's function formalism, which work in real space, instead of momentum space \cite{gennes_superconductivity_1999}.
Irrespective of the mathematical tools used, the complex order parameter $\Delta(\mathbf{r})$ must be calculated \emph{self-consistently} for current to be conserved \cite{gennes_superconductivity_1999,sols_crossover_1994,sonin_ballistic_2021}.

The issue of current conservation in BCS/BdG was pointed out specifically for JJs by Sols and Ferrer in Ref.~\cite{sols_crossover_1994}: if the phase in the leads is kept constant, yet there is a Josephson current in the junction, the solution cannot conserve current.
A resolution to the issue was left open in the scope of microscopic theory, and instead an analysis based on Ginzburg-Landau theory was pursued, in which the underlying microscopic quasiparticle spectrum is neglected and the issue of self-consistency is avoided entirely.
In a JJ, the self-consistency of $\Delta$ ensures the appropriate conversion of current carried by Andreev states to condensate current, as the self-consistent condition imposes that the Cooper pair condensate is in equilibrium with the unpaired electrons.
However, it is common practice to ignore the self-consistent condition, and assume fixed, constant phases in the superconducting leads, leading to the breaking of current conservation.

Within BdG theory, a prototypical model of a JJ is the superconductor--normal metal--superconductor (SNS) junction.
The SNS junction can easily be captured by including a region where the order parameter is required to vanish, which represents the normal metal.
This was already realized long ago, when the foundational work on SNS junctions was first performed \cite{kulik_macroscopic_1969,ishii1970Josephson,bardeen1972Josephson,svidzinsky1973Concerning,kulik1978Josephson}. 
These articles proposed various ways to calculate the Josephson current through an SNS junction, yet they all discarded the self-consistent condition, and modeled the superconducting order parameter as a \emph{square well} \cite{haberkorn1978Theoretical,arnold1985Superconducting,thuneberg2024Squarewell}.
As a result, the systems under study did not conserve current.
This lack of current conservation is not an issue in setups where the current density in the leads is small enough to be negligible; such as in constriction junctions, or tunnel junctions.
However, when this is not the case, the effects of current conservation on the CPR can be significant.


There exist in the literature several articles which perform self-consistent calculations, though the calculations are computationally demanding.
The methods used range from semi-analytical solutions of the BdG equations in slabs, where the boundary conditions are satisfied numerically \cite{kim2005Currentphase,spuntarelli_josephson_2007,spuntarelli_solution_2010}; semiclassical Green's function methods \cite{kupriyanov1992Effect,ruzhickiy2023Contribution}; and exact diagonalization \cite{covaci2006Proximity,black-schaffer_self-consistent_2008}.
These methods typically either find themselves in the semi-classical or Andreev limit ($\Delta_0\ll\mu$), or cannot easily be extended to more complex geometries, systems with impurities and wider parameter ranges.
In recent studies, various treatments of SNS JJs have been discussed \cite{sonin_ballistic_2021, thuneberg2024Squarewell}, culminating in recent works where Galilean transformations are applied to an SNS system to obtain current-conserving JJ solutions \cite{sonin2024Andreev,sonin2025theory}, even without self-consistency.
Thus, a general framework for fully current-conserving self-consistent solutions of the BdG equations has not yet been established.

In this article, we investigate the effects of current conservation on the CPR of a quasi-1D ballistic SNS junction at zero temperature, using the BdG formalism that we introduce in Sections \ref{sec:self-consistency_and_current} and \ref{sec:SNS_junction}.
In Sec. \ref{sec:CC_SC_scheme}, we introduce a general numerical method to solve the BdG equations self-consistently with current conservation in an SNS setup, which can be extended to model other junction types, as well as junctions in higher dimensions and at finite temperatures.
The method consists of self-consistently imposing a phase gradient in the superconducting leads, such that the current in the junction and the leads are equal.
In Section \ref{sec:analysis_current} we compare the CPRs obtained by our current conserving method with the results that we obtain by the standard method without current conservation. 
Then in Section \ref{sec:len_variation} we use our model to study the impact of junction length on the CPR and we further investigate the importance of current conservation by comparing our results to those obtained by standard methods \cite{thuneberg2024Squarewell} and recent analytical work that incorporates phase gradients \cite{sonin2025theory}. 
Similarly, in Section \ref{sec:gate_voltage} we employ our model to study the influence of a gate under the junction on the CPR. 
We find striking similarities between the CPRs obtained by our model and those experimentally observed in InAs nanowires \cite{spanton2017Current}. 
This ability of our model to reproduce the experimental data is reflected in multiple features of the CPR and its gate voltage dependence that are not captured by standard non-current-conserving methods.

\section{Self-Consistency and Current} \label{sec:self-consistency_and_current}

We describe our system using the BdG formalism \cite{gennes_superconductivity_1999,zhu_bogoliubov-gennes_2016}; the Bogoliubov equations are,
\begin{equation}\label{eq:BdG}
     \mathcal{H}_\mathrm{BdG} \begin{pmatrix}
         u_n\\ v_n
     \end{pmatrix}=E_n\begin{pmatrix}
         u_n\\v_n
     \end{pmatrix}
\end{equation}
where 
\begin{equation}
    \mathcal{H}_\mathrm{BdG} = \begin{pmatrix}
         H_0 - \mu & \Delta(\mathbf{r})\\
         \Delta^*(\mathbf{r}) & \mu -H_0
     \end{pmatrix}, \quad H_0 = -\frac{\hbar^2}{2m}\nabla^2 + V(\mathbf{r}),
\end{equation}
with $\mathcal{H}_\mathrm{BdG}$ the Bogoliubov-de Gennes Hamiltonian, $H_0$ the Hamiltonian for a quasi-free electron gas with effective mass $m$ and potential $V(\mathbf{r})$, $\mu$ the chemical potential, and $\Delta(\mathbf{r})=|\Delta(\mathbf{r})|e^{\imu\varphi(\mathbf{r})}$ the superconducting complex order parameter.
The self-consistency condition on $\Delta$, expressed in terms of the wavefunctions $u$ and $v$ is,
\begin{equation}\label{eq:delta_self_consistent_uv}
   \Delta(\mathbf{r})= U(\mathbf{r})\sum_{\mathclap{n}} u_n(\mathbf{r})v^*_{n}(\mathbf{r}) (1-2f_n)
\end{equation}
where $f_n=f(E_n) = 1/[\exp(\beta E_n) + 1]$ is the Fermi-Dirac distribution at inverse temperature $\beta$, $U(\mathbf{r})$ is the pairing energy and $n$ enumerates positive-energy solutions ($E_n > 0$).
\begin{widetext}
The charge and current density operators are given in terms of the Bogoliubov operators $\gamma_{n\sigma}$,
\begin{equation}
        \bm{\rho}(\mathbf{r}) = -e\sum_{\sigma n}|v_n|^2 - e\sum_{\sigma n m}\Big\{(u^*_nu_m-v_mv^*_n)\gamma^\dagger_{n\sigma}\gamma_{m\sigma}+\sigma u_n^*v^*_n\gamma^\dagger_{n\sigma}\gamma_{m-\sigma}^\dagger+ \sigma v_nu_m\gamma_{n-\sigma}\gamma_{m\sigma}\Big\},
\end{equation}
and
\begin{equation}
     \begin{aligned}\mathbf{j}(\mathbf{r})= -\frac{e \hbar}{2m\imu }\sum_{\sigma n}\left(v^*_n\nabla v_n - v_{n} \nabla v^*_{n}\right) -
     \frac{e \hbar}{2m\imu } \sum_{\mathclap{\sigma\, n\, m}} &\Big\{\left( u^*_{n}\nabla u_m - u_{m}\nabla u^*_{n} \right)\gamma_{n\sigma}^\dagger\gamma_{m\sigma} 
     + \left(v^*_{m} \nabla v_{n} - v_n\nabla v^*_m\right)\gamma^\dagger_{m-\sigma}\gamma_{n-\sigma} \\
    &+\sigma\left( u^*_{n}\nabla v^*_m -  v^*_{m}\nabla u^*_{n} \right)\gamma_{n\sigma}^\dagger\gamma^\dagger_{m-\sigma} +\sigma \left( v_n\nabla u_m -  u_{m} \nabla v_{n}\right)\gamma_{n-\sigma}\gamma_{m\sigma}\Big\},
    \end{aligned}
\end{equation}
\end{widetext}
where $\sigma=\pm1$ is a spin index and $e = |e|$ the fundamental charge.
We present the charge density and current operators in their full operator form, as these quantities are important to the current conservation that is central to our approach in this work.
One can clearly distinguish between the condensate terms (the first contributions in each equation) and the quasiparticle excitation terms.

The standard expression for the current \cite{blonder1982Transitionb} is recovered as the expectation value of the current operator,
\begin{equation}\label{eq:avg_current}
    \big\langle\,\mathbf{j}(\mathbf{r})\big\rangle = -\frac{2e\hbar}{m} \sum_{n}\mathrm{Im}\Big\{ u^*_{n}\nabla u_n f_n - v^*_{n} \nabla v_{n}(1-f_n)\Big\},
\end{equation}
where the factor of two comes from the sum over the spin index.

From the operator expressions, it can be shown that the charge continuity equation has a source term,
\begin{equation}\label{eq:current_continuity_equation}
	\left\langle \pdv{\bm{\rho}}{t} \right\rangle + \langle\div\mathbf{j}\rangle = -\frac{2e}{\hbar}\sum_{n}\mathrm{Im}\{2\Delta^*u_nv_n^*(2f_n-1)\}
\end{equation}
which is ensured to be zero if $\Delta(\mathbf{r})$ is self-consistent \cite{sols_crossover_1994}.

\section{The SNS junction}\label{sec:SNS_junction}
\begin{figure}
    \centering
    \includegraphics[width=\linewidth]{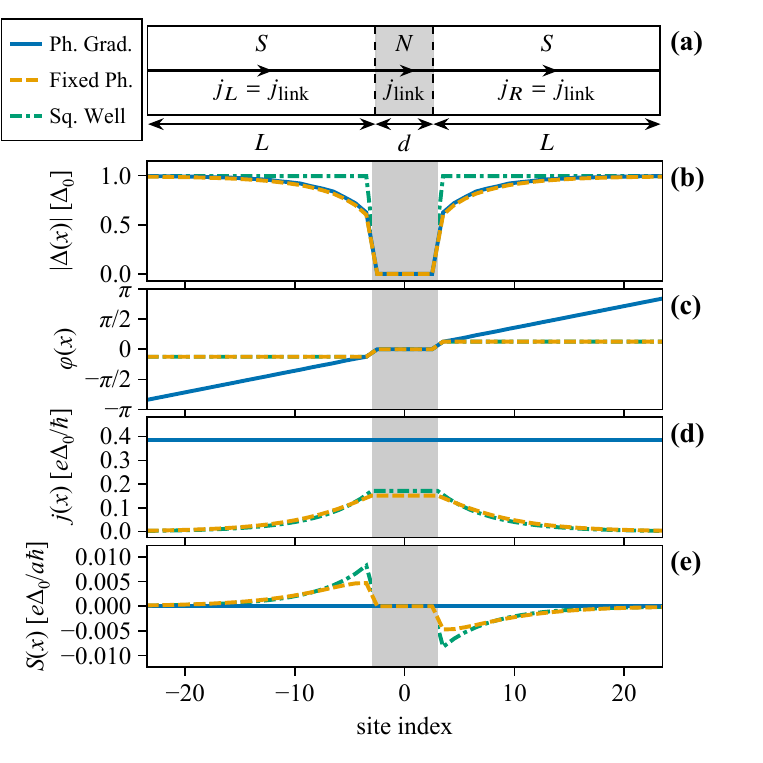}
    \caption{\textbf{(a)} Schematic of a Josephson Junction with $L$ and $d$ the lengths of the superconducting leads and normal (insulating or metallic) weak link, respectively. The transverse extent of leads and weak link are identical, which implies that the current density should match.
    \textbf{(b)--(e)} Solid blue lines represent the phase-gradient-matching solution; dashed orange lines represent the fixed-phase method, and the dash-dotted green lines the square-well model for $\Delta$.
    The phase drop over the junction is $\delta\varphi=\pi/4$.
    \textbf{(b)} Absolute value of the order parameter $|\Delta(x)|$.
    \textbf{(c)} Phase profile of the order parameter $\varphi(x)$.
    \textbf{(d)} Current $j(x)$ throughout the system.
    \textbf{(e)} Current source $S(x)$ (the rhs of Eq.~\ref{eq:current_continuity_equation}).}
    \label{fig:SNS_CC}
\end{figure}

We model our SNS junction as a quasi-1D system with a normal metal region separating two superconducting leads.
Our current-conserving modeling approach can be extended to 2D or 3D,
but numerically this makes the self-consistent iteration more computationally intensive.
Therefore we restrict the scope of this article to one dimension.
The conventional approach to model an SNS junction in BdG theory is to define an ansatz for the form of $\Delta$ as
\begin{equation}\label{eq:delta_step}
    \Delta(x) = \begin{cases}
        \Delta_0 e^{\imu \varphi_L} & x \leq -d/2\\
        0 &-d/2< x < d/2\\
        \Delta_0e^{\imu\varphi_R} & x \geq d/2
    \end{cases}
\end{equation}
where $d$ is the length of the junction, as schematically shown in Fig.~\ref{fig:SNS_CC}.
This ansatz may then be used in Eq.~\ref{eq:BdG} to calculate the eigenfunctions $(u_n,v_n)$ of $\mathcal{H}_\mathrm{BdG}$, which can be used to recalculate $\Delta$ (Eq.~\ref{eq:delta_self_consistent_uv}).
This is referred to as \emph{standard iteration} in Fig.~\ref{fig:cycle_diagram}.
As noted in Ref.~\cite{riedel_critical_1996}, under the standard iteration procedure, with nonzero initial phase drop $\delta\varphi=\varphi_R-\varphi_L$, the phase of the self-consistent order parameter $\Delta$ tends to a linear profile, which corresponds to a constant current in a bulk superconductor.
This tendency to linearity is frustrated if the phase is kept constant at the outer boundaries of the leads, and as a result the phase drop over the junction will decrease as the phase profile relaxes to an approximately linear profile.

In response to this, it is common practice to fix the phase in the leads and let only the magnitude of the complex order parameter $\Delta$ relax self-consistently under the conditions imposed on the phase profile.
We refer to this method as the \emph{fixed-phase} method (cf. Fig.~\ref{fig:cycle_diagram}).
In this model, the phase profile in the leads is constant (i.e. zero supercurrent), yet there \emph{is} a Josephson current, so the solution cannot be truly self-consistent, nor can it be current-conserving.

A current-conserving solution must incorporate a phase gradient in the leads. 
However, \emph{without} fixed phases in the leads, the definition of the phase drop becomes a subtle issue.
The phase drop over the junction $\big(\varphi(d/2)-\varphi(-d/2)\big)$ may be significantly different from the phase drop over the entire system $\big(\varphi(d/2+L)-\varphi(-d/2-L)\big)$ \cite{black-schaffer_self-consistent_2008}, where $L$ is the length of the leads (cf. Fig.~\ref{fig:SNS_CC} \textbf{(a)}).
From here on, we take the phase drop over the junction to denote the difference in phase immediately right and left of the junction: $\delta\varphi \equiv \varphi(d/2)-\varphi(-d/2)$.

\section{Current-conserving self-consistency scheme}\label{sec:CC_SC_scheme}
In a current-conserving junction, the leads carry the same net current as the junction.
That implies that in the leads, $\Delta$ should obtain a phase gradient $q$, proportional to the condensate motion, and matching the current in the junction due to the phase drop $\delta\varphi$.
Recognizing this, we propose a new ansatz for the phase profile and a modification of the self-consistent iteration to update $q$ to match the junction current.
The suggested ansatz incorporates a constant phase gradient $|\nabla\varphi| = q$ in the leads,
\begin{equation}\label{eq:delta_cc}
    \Delta(x) =\begin{cases}
        \Delta_0 e^{-\imu\delta\varphi/2} e^{\imu q(x+d/2)} & x\leq -d/2\\
        0 &-d/2 < x < d/2\\
        \Delta_0e^{\imu\delta\varphi/2}e^{\imu q(x-d/2)} & x \geq d/2
    \end{cases}
\end{equation}
where $q$ should ensure that the superconducting current in the leads matches the Josephson current through the weak link, and is determined self-consistently.

It is \emph{a priori} not clear what value $q$ should take. 
The relation between $q$ and the current is
\begin{equation}
    j_\mathrm{bulk}(x)  = \frac{e\hbar}{m}q\sum_n\left[|u_n(x)|^2f_n + |v_n(x)|^2(1-f_n)\right] 
\end{equation}
for $x$ in the superconducting bulk of the leads, i.e., $|x|-d/2\gg\xi $, where $\xi = \hbar v_F/\Delta_0$ is the coherence length.
We can exploit this monotonic dependence of $j_\mathrm{bulk}$ on $q$ to match the currents in the weak link and in the leads.
By letting $\Delta$ relax self-consistently with a linear phase gradient $q$ applied as boundary conditions (see Appendix~\ref{app:algorithm} and Fig.~\ref{fig:cycle_diagram} for details), the self-consistent solution converges to a current-conserving solution.
We refer to this method as the \emph{phase-gradient} method.

Specifically, we implement two modifications in the BdG self-consistency cycle.
First, we apply additional boundary conditions on $\Delta(x)$ with an imposed phase gradient $q$ that can be tuned so that the current in the leads matches the junction current. 
Second, we use the vanishing of the source term, i.e., the right-hand side of Eq.~\ref{eq:current_continuity_equation}, as a halting criterion for the algorithm, in addition to the absolute and relative tolerances.
This imposes a direct requirement of current conservation on the self-consistent iteration.
We apply the phase gradient $q$ in a region of $N_B$ sites extending from each boundary, with sources and sinks appearing there being considered outside of the system (see Appendix~\ref{app:numerical-solution} for details).
This approach ensures that the fundamental requirement of current conservation is satisfied throughout the system.

In Fig.~\ref{fig:SNS_CC}, we compare the fixed-phase and square-well methods to the proposed phase-gradient method.
Fig.~\ref{fig:SNS_CC} \textbf{(b)} shows $|\Delta(x)|$ for each method, showing almost no difference between the fixed-phase and phase-gradient methods.
The phase profiles, shown in \textbf{(c)}, are significantly different from each other, with the phase-gradient method, of course, introducing a gradient in the leads.
The current for the fixed-phase and square-well methods is localized around the junction and is manifestly not conserved (Fig.~\ref{fig:SNS_CC} \textbf{(d)}), as confirmed by the source term (Fig.~\ref{fig:SNS_CC} \textbf{(e)}) which is nonzero near the junction.
The phase-gradient solution displays a constant current, and a vanishing source term, as expected from a current-conserving solution.

\section{Analysis of the current-carrying state}\label{sec:analysis_current}
\begin{figure*}
    \centering
    \includegraphics[width=\linewidth]{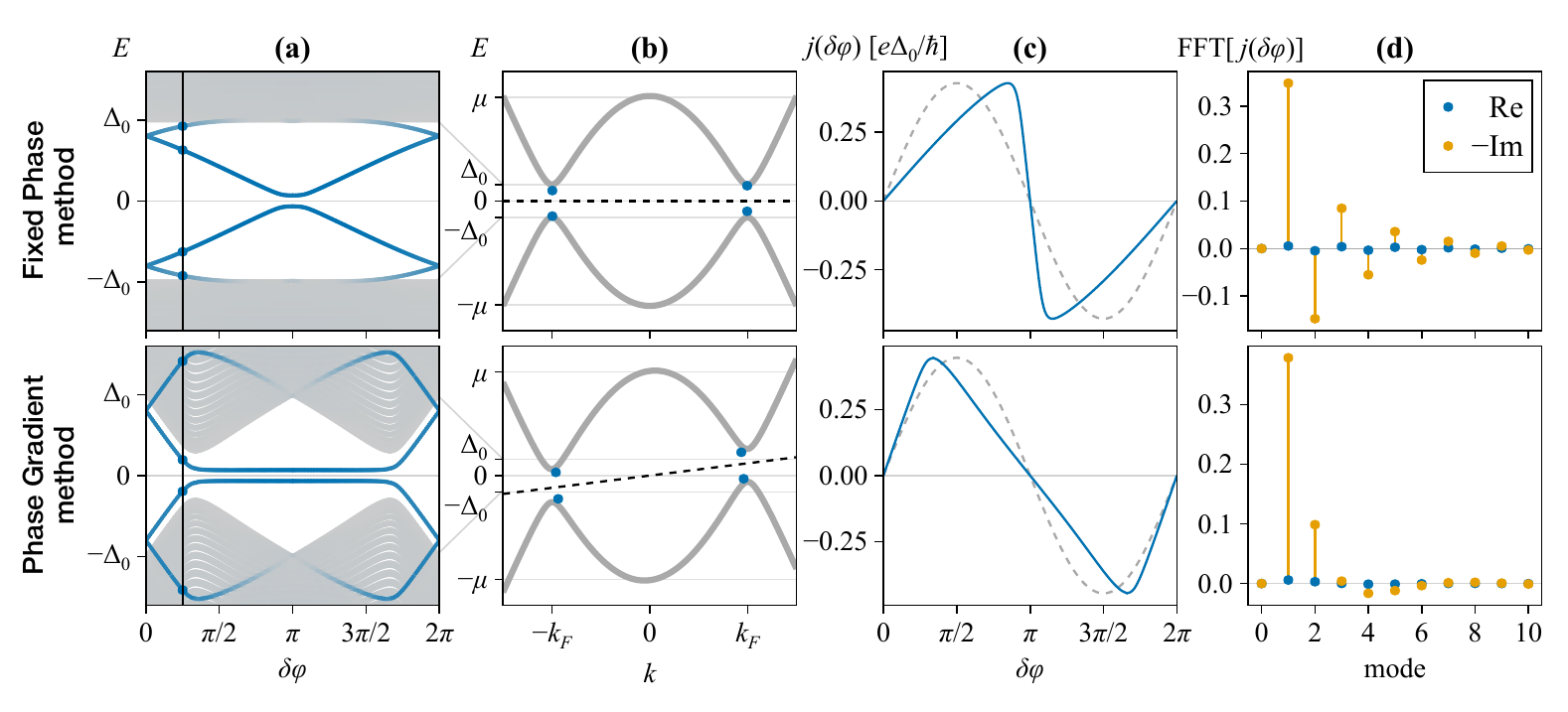}
    \caption{
    BdG spectrum, dispersion, CPR, and Fourier components of the CPR, as obtained with the (top row) fixed-phase and (bottom row) phase-gradient methods (see text for details). 
    \textbf{(a)} The BdG spectrum for varying phase drops over the junction.
    The Andreev states are identified by their localization within the normal metal, and are colored gray--blue for low and high degree of localization respectively. The Andreev states are blue.
    \textbf{(b)} The dispersion relation (Eq.~\ref{eq:dispersion}) with the Andreev bound states displayed as blue dots. The plots clearly show the raising and lowering of the Andreev states at $\pm k_F$. The dispersion is drawn for $\delta\varphi = \pi/4$ as indicated  by the vertical line in \textbf{(a)}.
    \textbf{(c)} The CPR with the dashed line showing a sinusoidal CPR as reference. 
    \textbf{(d)} Fourier components of the CPRs. 
    The phase gradient in the leads strongly affects the Josephson harmonics.
    The length of the junction is $d \approx 2.2 \, \xi$.
    }
    \label{fig:spectrum_cc}
\end{figure*}

We proceed in our study by analyzing the BdG spectrum of our SNS system (Fig.~\ref{fig:spectrum_cc} \textbf{(a-b)}) and compare between fixed-phase and phase-gradient approaches.
We then evaluate the junction currents and address the differences between the CPRs obtained by the fixed-phase and phase-gradient methods.

In the fixed-phase method, the current is predominantly carried by the Andreev bound states (ABS), with their contribution being proportional to the derivative of their energies to the phase difference over the junction \cite{thuneberg2024Squarewell}
\begin{equation}\label{eq:ABS_current}
    j_{\mathrm{ABS}} = \frac{2e}{\hbar} \sum_n \dv{E_n}{(\delta\varphi)}f(E_n)
\end{equation}
where $n$ enumerates the Andreev levels.
In the Andreev limit ($\Delta_0\ll\mu$), for perfectly transparent junctions, analytic expressions for the ABS energies are known, and may be used to calculate CPRs for the fixed-phase method.
In Fig.~\ref{fig:spectrum_cc} \textbf{(a)}, top row, the BdG spectrum is plotted for the fixed-phase method.
The spectrum was obtained numerically for $\mu/\Delta _0\approx5$.
Note that this ratio is not deep in the Andreev limit such that an anticrossing with small splitting is produced at $\delta\varphi=\pi$ instead of a level crossing (see Appendix \ref{app:transparency}). 
The CPR (Fig.~\ref{fig:spectrum_cc} \textbf{(c)}) indeed seems qualitatively proportional to the phase-derivative of the ABS.
Comparing to the bottom row of Fig.~\ref{fig:spectrum_cc}, showing the corresponding plots for the phase-gradient method, clearly the same statement cannot be made for the Andreev states in the bottom row of Fig.~\ref{fig:spectrum_cc} \textbf{(a)}.
It is clear that the state of the superconducting leads affects the BdG spectrum strongly, thereby invalidating the approach based on Eq.~\ref{eq:ABS_current}.

When the condensate is moving at a group velocity $v_q = \hbar q/m$, i.e. $\Delta \to \Delta e^{\imu q x}$, the quasiparticle excitations will be excited with respect to the moving condensate of pairs $(k+q,\,-k+q)$.
That is, the quasiparticles are \emph{Doppler-shifted} by $q$,
\begin{equation}
    \begin{pmatrix}
        u(x)\\ v(x)
    \end{pmatrix} \to \begin{pmatrix}
        u(x)e^{\imu q x/2} \\ v(x)e^{-\imu q x/2}
    \end{pmatrix}.
\end{equation}
This changes the dispersion relation of the superconducting leads (Fig.~\ref{fig:spectrum_cc} \textbf{(b)}), which becomes \cite{gennes_superconductivity_1999}:
\begin{equation}\label{eq:dispersion}
    E_{{k}} = \frac{\hbar^{2}}{m} k \cdot q \pm \sqrt{\left( \frac{\hbar^{2}}{2m} ({k}^{2} + {q}^{2}) - \mu \right)^{2} + |\Delta|^{2}} 
\end{equation}
This lowers (increases) the energy of the states at $-k_F$ ($+k_F$) proportionally to $q$ (as shown in Fig.~\ref{fig:spectrum_cc} \textbf{(b)}), eventually closing the gap at $q \approx \xi^{-1}$, which is when the superconducting leads reach their depairing current \cite{bagwell1994Critical}. 
This energy shift has profound consequences for the BdG spectrum. 
As Fig.~\ref{fig:spectrum_cc} \textbf{(a)} demonstrates, the Andreev bound states in the phase-gradient approach look qualitatively very different from those in fixed-phase calculations.
The Andreev states (blue in Figs.~\ref{fig:spectrum_cc} \textbf{(a)}-\textbf{(b)}) are shifted up and down together with the bulk states, substantially modifying their energies and current-carrying properties.

The current-conservation requirement also significantly affects the CPR (see Fig.~\ref{fig:spectrum_cc} \textbf{(c)}), exhibiting backward skewness where the phase angle where the maximal current is reached $\varphi_\mathrm{max} < \pi/2$, rather than the forward skewness ($\varphi_\mathrm{max}>\pi/2$) typically seen in conventional models of SNS junctions. 
This reversal aligns with predictions from Sols \& Ferrer using Ginzburg-Landau theory \cite{sols_crossover_1994}, and some self-consistent BdG calculations \cite{freericks2002Optimizing,kim2005Currentphase,golubov_current-phase_2004,spuntarelli_josephson_2007}, though it contradicts standard SNS CPR theory \cite{haberkorn1978Theoretical,arnold1985Superconducting,thuneberg2024Squarewell}.
Generally, the phase-gradient method lessens the forward skewness of the CPR, compared to the fixed-phase CPR (cf. Fig.~\ref{fig:VG_CPRs}).
Fourier analysis reveals that the sign of the second harmonic is flipped (same sign as first harmonic) in backward-skewed CPRs, as shown in Fig.~\ref{fig:spectrum_cc} \textbf{(d)}.
%

Note that for superconductor--insulator--superconductor (SIS) junctions, the currents carried by the junction are of much lower magnitude.
Therefore, $q$ is also significantly smaller than in the SNS junction, and the modification of the CPR becomes negligible.
One can model SIS junctions in the framework presented here by raising a potential $V(x)$ inside the normal metal, effectively turning the normal metal into an insulator; see also Sec.~\ref{sec:gate_voltage}.

\section{Junction Length Variation}\label{sec:len_variation}
\begin{figure*}
    \centering
    \includegraphics[width=\linewidth]{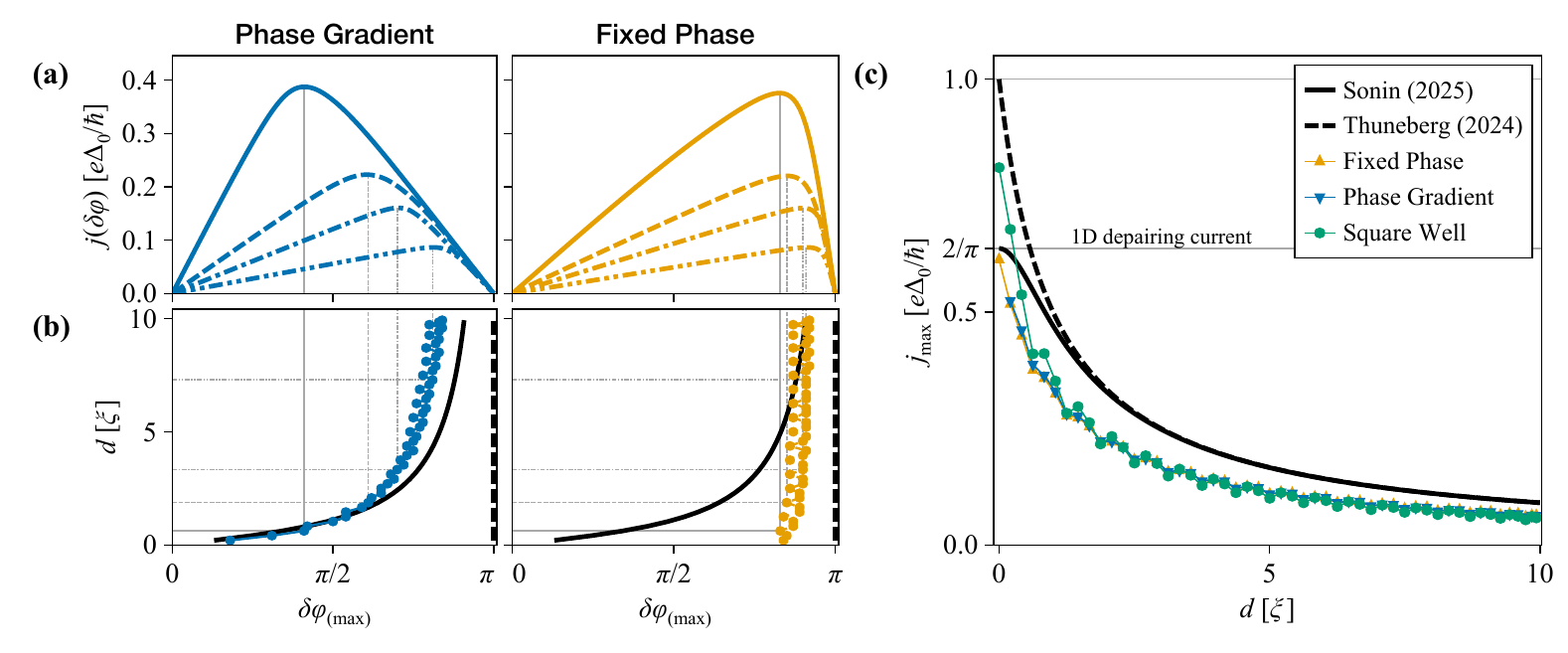}
    \caption{CPRs for junctions transitioning from the short junction ($d<\xi$) to the long junction regime ($d>\xi$); $\xi \approx 10$ sites.
            \textbf{(a)} CPRs for the phase-gradient and fixed-phase methods. 
            \textbf{(b)} The evolution of the CPR maximum $\delta\varphi_{\mathrm{max}}$ with junction length. The colored dots represent numerical results, and the solid and dashed black lines derive from Refs.~\cite{sonin2025theory} and \cite{thuneberg2024Squarewell} respectively.
            \textbf{(c)} The maximal values reached by the CPRs plotted against the junction length. The phase-gradient and fixed-phase methods agree remarkably well despite their difference in shape.
    }
    \label{fig:L_variation}
\end{figure*}
We can gain some insight into the shape and properties of the current-conserving CPR by calculating the CPR for different junction lengths, from short ($d\ll\xi$) junctions to long ($d\gg\xi$) junctions.
Variation of junction length is known to influence transport properties strongly \cite{levajac2023Impact}, with the maximum amplitude typically varying as $j_{\mathrm{max}}\sim 1/d$ \cite{thuneberg2024Squarewell}.
The resulting CPRs for both the phase-gradient and fixed-phase methods are shown in Fig.~\ref{fig:L_variation} \textbf{(a)}.
The phase drop at which the maximal current is reached, $\delta\varphi_{\mathrm{max}}$ is significantly different in the phase-gradient CPR compared to the fixed-phase CPR (Fig.~\ref{fig:L_variation} \textbf{(b)}).
The fixed-phase CPRs are always forward-skewed, while for the phase-gradient CPRs, $\delta\varphi_{\mathrm{max}}$ increases from below $\pi/2$ (backward-skewed) for short junctions to above $\pi/2$ (forward skewed) for long junctions.
The maximal value of the current $j_{\mathrm{max}}$, remarkably remains in good agreement between the two approaches (Fig.~\ref{fig:L_variation} \textbf{(c)}).

We can compare our numerical results to a recent work which predicts current-conserving CPRs for ballistic planar SNS junctions at $T=0$ (Sonin \cite{sonin2025theory}), and also to the fixed-phase square-well theory at $T=0$, as derived in Refs.~\cite{ishii1970Josephson,svidzinsky1973Concerning,kulik1978Josephson} and recently summarized by Thuneberg \cite{thuneberg2024Squarewell}.
Sonin's method calculates CPRs by applying Galilean boosts to the SNS system to obtain solutions where the current in the leads is equal to that in the normal metal.
Thuneberg's method relies on finding the BdG spectrum for a square-well profile for the order parameter, and summing the Andreev and continuum contributions.
Both methods make the Andreev approximation $\Delta_0\ll\mu$, which our method does not.

The properties of the CPRs derived from Sonin and Thuneberg's methods are indicated in Fig.~\ref{fig:L_variation} by solid and dashed lines respectively.
Sonin's prediction for $\delta\varphi_{\mathrm{max}}$ (Fig.~\ref{fig:L_variation} (\textbf{b})) agrees well qualitatively with our numerical results, producing backward-skewed CPRs for short junctions, and becoming increasingly forward-skewed as the junction length increases.
This behavior is not captured at all by the fixed-phase method, as this method only produces forward-skewed CPRs as $d$ increases.
Both Sonin and Thuneberg's prediction for $\delta\varphi_{\mathrm{max}}$ is $\delta\varphi_{\mathrm{max}}=\pi$ as $d\to\infty$.
Our numerics result in lower $\delta\varphi_{\mathrm{max}}$, as we are not in the limit $\Delta_0\ll\mu$, and therefore normal scattering is not completely suppressed, and the junction is not perfectly transparent (see Appendix \ref{app:transparency}).

The predictions for the $j_{\mathrm{max}}$ of the junction differ most at the limiting case $d=0$.
The critical current of a 1D superconductor (with no junction) was reported by Bagwell to be $2e\Delta_0/(\pi\hbar)$ \cite{bagwell1994Critical}, and is the maximal Josephson current predicted by Sonin through a junction of length $d\to0$, at phase drop zero.
Thuneberg's maximal current exceeds this current, as the predicted maximal current is $e\Delta/\hbar$, at phase drop $\pi$.
We have numerically confirmed that the critical current of our 1D superconducting system (with no junction) is $2e\Delta_0/(\pi\hbar)$.
Any current higher than that value cannot be carried by the leads without closing the gap (cf. Eq.~\ref{eq:dispersion}).
Interestingly, this critical current limit is also obtained with the fixed-phase method, even though there is no condensate current far in the leads, as there is no phase gradient. 
In this case, the critical current limit is guaranteed through the self-consistency of the modulus of $\Delta$ near the junction (cf. Fig.~\ref{fig:SNS_CC} \textbf{(b)}), which also represents a conversion of Andreev current into condensate current. 
As there is no current conservation with the fixed-phase method, this condensate current is not carried further into the leads.
However, if we assume a pure square well, without \emph{any} self-consistency, we observe that the maximal $j_{\mathrm{max}}$ does tend to $e\Delta_0/\hbar$.
Of course, both the fixed phase and the square well are idealizations in a purely 1D system.
If the leads can be understood to be higher dimensional reservoirs with a cross section much greater than that of the quasi-1D normal metal, the depairing current of the leads will be many times greater than $e\Delta_0/\hbar$, and the current is limited by the junction.
However, if the leads are also quasi-one dimensional, the current through the junction will be fundamentally limited by the current that can be transported by the leads.
\begin{table}[h]
    \centering
    \begin{tabular}{c|cc}
        {} & $j_{\mathrm{max}}$ & $\delta\varphi_{\mathrm{max}}$\\
        \hline
        Sonin (\citeyear{sonin2025theory}) & $\frac{2e\Delta_0}{\pi\hbar}$ & $0$\\
        Thuneberg (\citeyear{thuneberg2024Squarewell}) & $\frac{e\Delta_0}{\hbar}$ & $\pi$\\
        
    \end{tabular}
    \caption{Comparison between Sonin and Thuneberg's predictions for the CPR of a $d=0$ junction at zero temperature.}
\end{table}

For long ballistic junctions, $d\gg\xi$, Sonin and Thuneberg's models for $j_{\mathrm{max}}$ agree perfectly.
So do our numerical results, although our reported $j_{\mathrm{max}}$ is slightly lower due to the imperfect transparency of the simulated junctions (see Appendix \ref{app:transparency}).
In the limit $d\to\infty$, all models discussed here agree that $j_{\mathrm{max}}\to0$ and $\delta\varphi_{\mathrm{max}}\to\pi$.

We also note that the BdG spectrum for long junctions is less perturbed by the bulk states lowering in energy than that of the short junction.
The lessened effect of the superconducting states in the leads on the states in the gap can be understood by considering the fact that the superconducting bulk states lower into the gap proportionally to the magnitude of $q$.
For longer junctions, the current is decreased, therefore so is $q$ and the CPRs look more like in the standard picture.

Note that, at finite temperatures, as $d$ increases, the current becomes exponentially suppressed on the order of the thermal dephasing length, $\xi_T = \hbar v_F / k_BT$ \cite{bagwell1994Critical}.
In the limit $d\to\infty$, the maximum current will exponentially approach 0, and the critical angle will approach $\pi/2$ \cite{eley2013Dependence}.

\section{Applying Gate Voltage}\label{sec:gate_voltage}
\begin{figure}
    \centering
    \includegraphics[width=\linewidth]{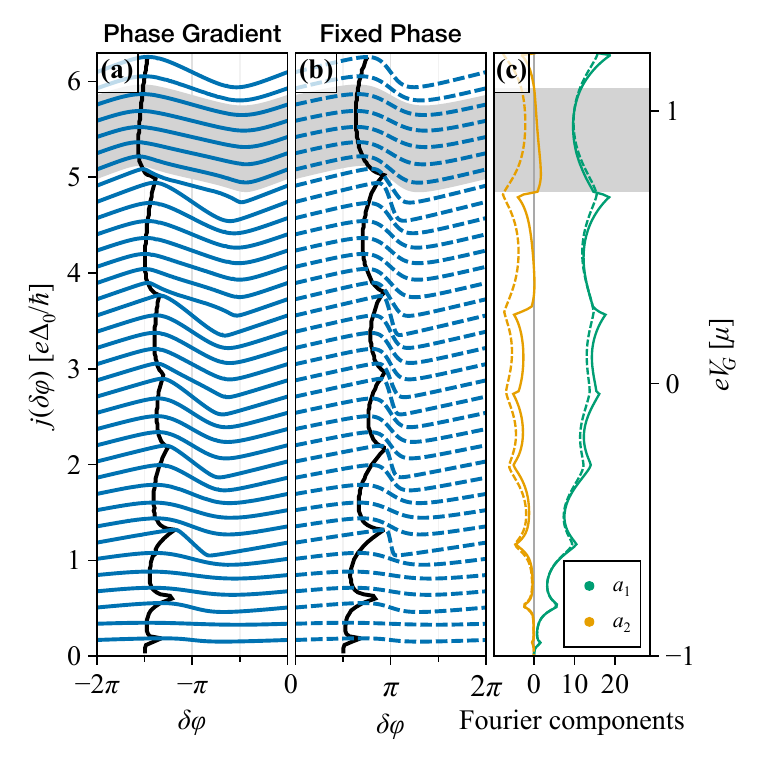}
    \caption{\textbf{(a)} CPRs of the phase-gradient method with varying gate voltages applied (offset according to $V_G$). 
    The black line tracks the maxima of the CPRs.
    The CPRs in the shaded region display backward skewness.
    \textbf{(b)} CPRs of the fixed-phase method. 
    All CPRs are forward-skewed.
    \textbf{(c)} First two Fourier components (imaginary parts), $-\mathrm{Im}[a_{1,2}]$, for the phase-gradient (solid) and fixed-phase (dashed) methods. 
    The shaded area marks the $V_G$-interval where $a_1$ and $a_2$ are of the same sign, indicating the backward skewness obtained by the phase-gradient method. 
    $eV_G = -1\mu$ corresponds to depletion and marks the crossover from the SNS to SIS regime. In calculating these CPRs, $d/\xi \approx 3$.}
    \label{fig:VG_CPRs}
\end{figure}
Our numerical implementation allows us to go beyond the Andreev approximation, but also beyond assumptions such as the equal Fermi energies of the superconductor and the normal metal. 
We make use of this flexibility to calculate the CPR while applying a potential barrier of varying strength inside the normal metal part: $V(x) = -eV_G$ if $|x|<d/2$ and zero otherwise.
This is analogous to applying a gate voltage to a gate under a nanowire junction.
The chemical potential $\mu$ is still kept constant throughout the system but, by applying a potential within the normal metal, one effectively modifies the electro-chemical potential to $\mu_{\mathrm{eff}}=\mu + eV_G$.
In this manner the number of available conduction electrons can be tuned.
The resulting CPRs are shown in Figs.~\ref{fig:VG_CPRs} \textbf{(a)} and \textbf{(b)} for the phase-gradient and fixed-phase methods respectively.
The evolution of the skewness is traced by the black line behind the CPRs, which connects the maxima of the offset CPRs.
In order to quantify the evolution of the shape of the CPRs, we extract the first two Fourier components of the CPR and plot their evolution with $V_G$ in Fig.~\ref{fig:VG_CPRs} \textbf{(c)}.
If the first two Fourier components have opposite sign, the CPR is forward-skewed, otherwise, the CPR is backward-skewed.

For the phase-gradient method, Fig.~\ref{fig:VG_CPRs} \textbf{(a)}, we see the appearance of backward skewed CPRs for a range of gate voltages around $eV_G \approx 1\mu$ (shaded region). 
For lower values of $V_G$, the CPR is skewed forward.
Comparing directly with the CPRs obtained from the fixed-phase method (Fig.~\ref{fig:VG_CPRs} \textbf{(b)}), we see that the shapes of the CPRs are significantly different.
In the fixed-phase method, there is no reversal of skewness, and \emph{only} forward skewed CPRs are obtained.
We can confirm this by looking at the first two Fourier components (shown in Fig.~\ref{fig:VG_CPRs} \textbf{(c)}) which have opposite sign for forward-skewed CPRs and the same sign otherwise.

Recent measurements of InAs nanowire junctions \cite{spanton2017Current} show similar qualitative features as the CPRs obtained using the phase-gradient method (Fig. \ref{fig:VG_CPRs} \textbf{(a)}). 
Most notable is the appearance of backward skewness in a certain regime of $V_G$, which cannot be reproduced using the fixed-phase methods, yet appear naturally for short junctions and for certain gate voltages when ensuring current conservation.

\section{Conclusion}
Current conservation is only guaranteed in a BdG system if the order parameter is treated fully self-consistently.
Disregard of this fact may have severe consequences in setups where the supercurrent density is not negligible.
We analyze the consequences of the proper implementation of self-consistency for the CPR in SNS junctions.
To do this, we propose and evaluate a current-conserving self-consistency scheme for the microscopic BdG modeling of a Josephson junction, by applying a phase gradient in the leads to match the current through the junction.
As a consequence of ensuring current conservation, we find significant modifications to the BdG spectrum and to the CPR, in cases where the current density in the leads is non-negligible.
The superconducting bulk states in the leads now also carry current and descend below $\Delta$, severely affecting the BdG spectrum.
When current densities are small compared to the critical current of the leads, the qualitative change is small, but in, e.g., nanowires this is not guaranteed to be the case.
We find our approach to be in good agreement with recent analytical work concerning the impact of junction length on the CPR, and show that variation of a gate voltage controlling the normal metal can lead to backward-skewed CPRs.
Backward skewness of the CPR is opposite to what is obtained with conventional non-current-conserving approaches.
In this way the phase-gradient approach opens the door to understanding the backward-skewed CPRs observed in InAs nanowires \cite{spanton2017Current}.
The phase-gradient method opens up new research avenues regarding the effect of current-carrying bulk states on the BdG spectrum and their respective contributions to the Josephson current.
The method naturally produces backward-skewed CPRs, and may thus be used to cast light on which factors determine the precise shape of the CPR.

\appendix

\section{Numerical solution of the BdG system}
\label{app:numerical-solution}
We aim to find solutions $(u_n,v_n)$ to the Bogoliubov equations, Eq.~\ref{eq:BdG}. 
These consist of two coupled time-independent Schr\"odinger-type equations, which form an eigenvalue problem.
We employ straightforward finite-difference discretization to obtain the Hamiltonian matrix, which we then diagonalize using standard routines such as those provided by the \texttt{LAPACK} library.

In this work, we consider a 1D system, let $\hbar=m=1$, and set the grid spacing $a = 1$, with 200 sites.
This means our effective coupling between sites is $t = 1/(2a^2) = 1/2$.
The dispersion relation is that of a tight-binding model, $E = \mu + 2t\big(1-\cos(ka)\big)$ where the constant $2t$ appears naturally as a result of the finite-difference discretization, and ensures that $\mu = 0$ corresponds to the empty band.

We set $\mu = t$ and aim for $|\Delta|$ to be approximately $t/5$, which we satisfy self-consistently by tuning the pairing energy to $U = 1.65t$.
This way, the coherence length spans approx. 10 sites.
The width of the normal metal $d$ was set to 10 sites in Figs.~\ref{fig:SNS_CC}--\ref{fig:spectrum_cc}, and 20 sites in Fig.~\ref{fig:VG_CPRs}.
All calculations were performed at $T=0$ such that $f_n = 0$ for all $n$ with $E_n>0$, and $f_n=1$ for $n$ with  $E_n<0$.
The temperature enters the calculation only in the self-consistency equation and in the current evaluation.

\begin{figure}
    \centering
    \includegraphics[width=\linewidth]{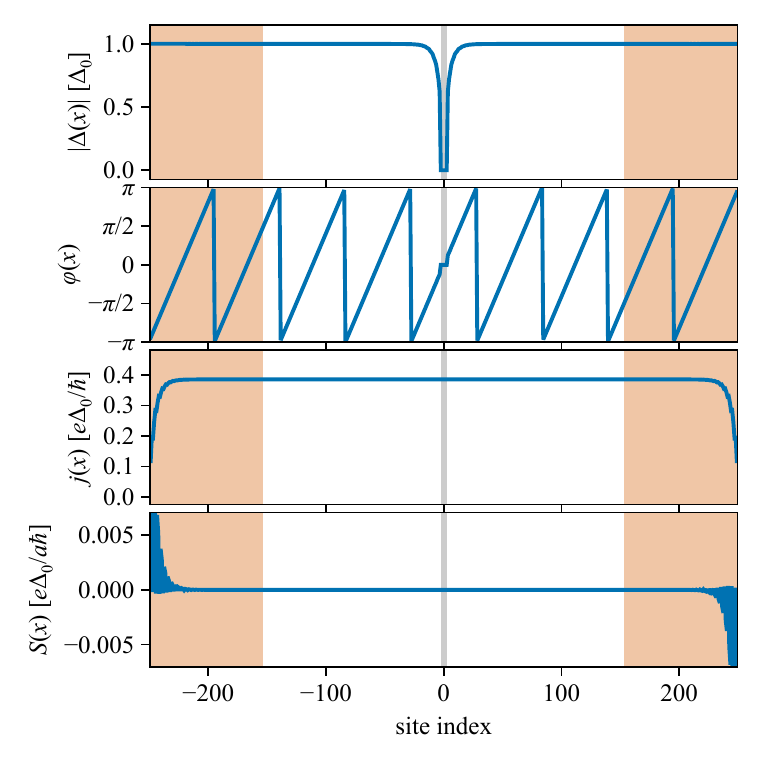}
    \caption{Reproduction of Fig.~\ref{fig:SNS_CC} with the boundary regions included (shaded orange).}
    \label{fig:boundary}
\end{figure}

The boundary conditions on $u$ and $v$ are Dirichlet zero boundary conditions.
Direct application of the self-consistency condition of Eq.~\ref{eq:delta_self_consistent_uv} would then lead to zero $\Delta$ at the boundaries too.
This is undesirable for our purposes, as we want to impose a fixed phase gradient also at the edges.
Therefore, we impose additional conditions on $\Delta$ at each self-consistent iteration: the phase is set to have a constant gradient $q$ for $N_B$ sites from the boundary, and the magnitude $|\Delta|$ is set to remain constant for $N_{B_a}$ sites from the boundary.
 $N_B$ and  $N_{B_a}$ need not be the same.
We interpret the zones under these additional constraints as being themselves part of the boundary, such that the central area where the self-consistent calculation is not constrained can be considered the true system under study (see Fig.~\ref{fig:boundary}).

\begin{figure*}
    \centering
    \includegraphics[]{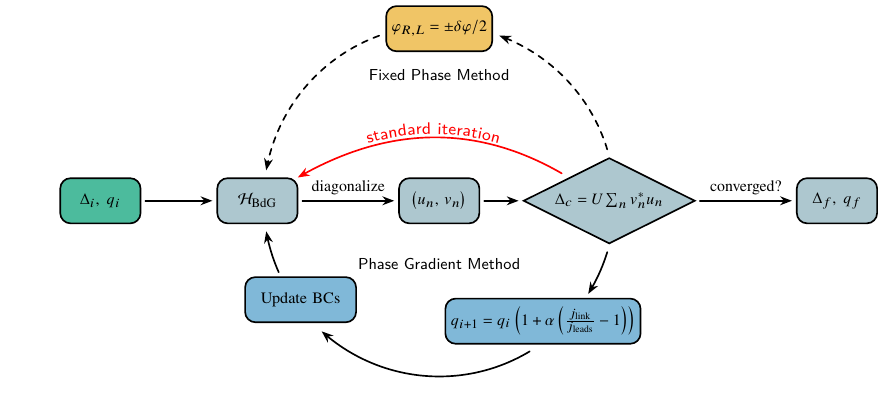}
    \caption{Diagram of the iterative self-consistent calculation. The upper part of the diagram shows the fixed-phase method, where the phase is kept fixed at every iteration. The lower part shows the phase-gradient method, where the boundary conditions are updated so the current in the leads matches the junction current. The convergence criteria also differ among methods.}
    \label{fig:cycle_diagram}
\end{figure*}

\section{Phase-Gradient algorithm}
\label{app:algorithm}

The self-consistency algorithm is schematically illustrated in Fig.~\ref{fig:cycle_diagram}, and starts from the ansatz in Eq.~\ref{eq:delta_cc} with an estimate for $\Delta$ and $q$. 
We assign a number of sites $N_B$ from each lattice end to represent the boundary, where the phase gradient will be set to $q$.
Then, the algorithm goes as follows
\begin{enumerate}
    \item \label{step:eigs}Calculate the eigenvalues and eigenvectors of $\mathcal{H}_\mathrm{BdG}$.
    \item Use the eigenvalues/vectors to evaluate the currents $j(0) \equiv j_\mathrm{link}$ and $j(N_B) \equiv j_\mathrm{leads}$.
    \item Update $q$ proportionally to the mismatch between $j_\mathrm{link}$ and $j_\mathrm{leads}$
    \[
    q_{i+1} = q_i  (1 + \alpha  (j_\mathrm{link}/j_\mathrm{leads} - 1)),
    \]
    where $\alpha$ is the mixing parameter.
    \item Update $\Delta(\mathbf{r})$ using the self-consistent equation (Eq.~\ref{eq:delta_self_consistent_uv}), and apply the updated phase gradient $q$ at the boundaries. Then update $\mathcal{H}_\mathrm{BdG}$ with the new $\Delta$  and return to step \ref{step:eigs}.
\end{enumerate}
This iteration can then be repeated until $j_\mathrm{link}$ and $j_\mathrm{leads}$ match up to the required tolerance and the source term in Eq.~\ref{eq:current_continuity_equation} is sufficiently small throughout the system.

There are several refinements which can be made to enable faster convergence.
In our calculations, we fix the phase drop over the weak link after every convergence by shifting the entire phase profile in the respective leads up or down.
We also set the phase gradient in the outer region $N_B$ to be exactly constant and equal to $q$, whereas in the region around the weak link, variations are allowed by means of the self-consistent relaxation.
Finally, we also set the magnitude of $\Delta$ to be constant at the boundaries (in the region $N_{B_a}$) and equal to the value of $|\Delta|$ in the interior of the leads.

The self-consistency scheme has two convergence criteria: \textbf{1)} the matching of the current in the leads and in the normal metal up to a specified tolerance; \textbf{2)} the vanishing of the source term in Eq.~\ref{eq:current_continuity_equation} up to a specified tolerance.
The second is a \emph{physical} requirement of self-consistency since, as was pointed out above, the source term vanishes if the system is self-consistent.
The tolerances implemented for the simulations shown were $|j_{\mathrm{link}}-j_{\mathrm{leads}}|/j_{\mathrm{link}} < 10^{-6}$ and $aS / j_{\mathrm{link}} < 10^{-4}$.
The self-consistent loop converges to within these tolerances in $\sim200$ iterations.

The number of iterations needed is large compared to self-consistent calculations without current conservation; indeed, the magnitude $|\Delta|$ converges much faster than the phase. 
This is partially due to the absence of a good estimate for the initial value of $q$ and the shape of the phase profile.
Additionally, the finite size of the system entails that there will always remain some frustration of the system as a result of these boundary conditions.
For arbitrarily large systems, the tolerances can be made arbitrarily low.

\section{Junction Transparency}
\label{app:transparency}
\begin{figure}
    \centering
    \includegraphics[width=\linewidth]{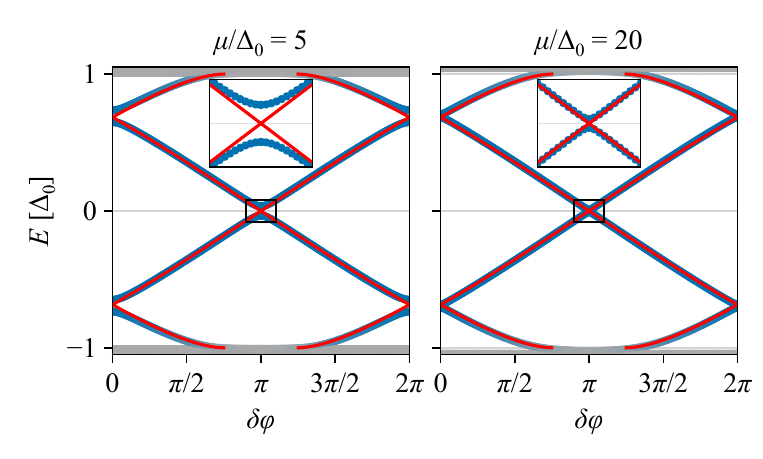}
    \caption{BdG spectrum in the square-well model. Left plot shows numerical BdG spectrum at $\mu/\Delta_0=5$; right plot shows $\mu/\Delta_0=20$. Red lines in both plots show the known analytic expression for $E_{\mathrm{ABS}}(\delta\varphi)$ in the square-well model, Eq.~\ref{eq:ABS}. The inset axes are enlargements of the anticrossings at $\delta\varphi=\pi$.}
    \label{fig:transparency}
\end{figure}
Despite there being no barriers at the SN interfaces, our simulated junctions are not perfectly transparent, as evidenced by the avoided crossing at $\delta\varphi=\pi$ in Fig.~\ref{fig:spectrum_cc} \textbf{(a)}; see also Fig.~\ref{fig:transparency}.
Since our system is not properly in the Andreev limit ($\Delta_0\ll\mu$), the Andreev reflection probability at the SN interfaces is not precisely 1 for sub-gap states.
This means that normal reflection is also possible, which causes a coupling between the positive and negative energy Andreev bands (which also have opposite momenta), with an avoided crossing as a result.
The maximal achievable ratio $\mu/\Delta_0$ is limited by the bandwidth of the tight-binding band obtained by the finite-difference discretization.
Discretizing with a finer lattice allows for higher values of $\mu/\Delta_0$, which further suppresses normal reflection and enhances Andreev reflection.
We can numerically confirm this in the square-well model for $\Delta(x)$, as shown in Fig.~\ref{fig:transparency}.
The discrete part of the spectrum of a ballistic SNS junction with perfect transparency may be found analytically (see e.g. Ref.~\cite{thuneberg2024Squarewell}).
The red curves in Fig.~\ref{fig:transparency} are given by
\begin{equation}\label{eq:ABS}
\begin{split}
    \delta\varphi(E) = \arccos\Bigg[ &-\left(1-2\frac{E^2}{\Delta_0^2}\right)\cos\frac{2Ed}{\Delta_0\xi} \\
    &+ 2\frac{E}{\Delta_0}\sqrt{1-\frac{E^2}{\Delta_0^2}}\sin\frac{2Ed}{\Delta_0\xi} \Bigg].
    \end{split}
\end{equation}
The Andreev limit $\Delta_0\ll\mu$ is well studied because of the simplifications of the equations when small fluctuations (of $\sim k_F$) are ignored.
The BdG equations may then be simplified to the Andreev equations, which involve only a single derivative and are therefore numerically and analytically easier to analyze.

\bibliography{ref}

\end{document}